\documentclass[useAMS,usenatbib]{biom}
\usepackage{hyperref} % causes a paper height warning, don't know why. The following commented line removes the warning, but changes the text alignment
\hypersetup{colorlinks, allcolors = blue, citecolor=blue, urlcolor=black}

\usepackage[dvipsnames]{xcolor} % colors math
\usepackage{amsmath, amsfonts, bbm}
\usepackage{booktabs}
\usepackage{mathtools}
\usepackage{pdfpages} % for tacking supplementary materiasl pdf at the end of the manuscript (for arxiv)

\newtheorem{proposition}{Proposition} % since biom.cls does not account for propositions

\def\bSig\mathbf{\Sigma}

\title[SD-Learning for Efficient Estimation of Individualized Treatment Rules]{Stabilized Direct Learning for Efficient Estimation of Individualized Treatment Rules}

\author{Kushal S. Shah$^{1,*}$\email{kushshah@live.unc.edu}, 
Haoda Fu$^{2,**}$\email{fu\_haoda@lilly.com}, and 
Michael R. Kosorok$^{1,***}$\email{kosorok@unc.edu} \\
$^{1}$Department of Biostatistics, University of North Carolina at Chapel Hill, NC 27599, USA \\
$^{2}$Eli Lilly and Company, Lilly Corporate Center, Indianapolis, IN 46285, USA}

\begin{document}

%  These options will count the number of pages and provide volume
%  and date information in the upper left hand corner of the top of the 
%  first page as in published papers.  The \pagerange command will only
%  work if you place the command \label{firstpage} near the beginning
%  of the document and \label{lastpage} at the end of the document, as we
%  have done in this template.

%  Again, putting a volume number and date is for your own amusement and
%  has no bearing on what actually happens to your paper!  

%\volume{64}
%\pubyear{2008}
%\artmonth{December}

%  The \doi command is where the DOI for your paper would be placed should it
%  be published.  Again, if you make one up and stick it here, it means 
%  nothing!

%\doi{10.1111/j.1541-0420.2005.00454.x}

%  This label and the label ``lastpage'' are used by the \pagerange
%  command above to give the page range for the article.  You may have 
%  to process the document twice to get this to match up with what you 
%  expect.  When using the referee option, this will not count the pages
%  with tables and figures.  

\label{firstpage}

%  put the summary for your paper here

\begin{abstract}
In recent years, the field of precision medicine has seen many advancements. Significant focus has been placed on creating algorithms to estimate individualized treatment rules (ITR), which map from patient covariates to the space of available treatments with the goal of maximizing patient outcome. Direct Learning (D-Learning) is a recent one-step method which estimates the ITR by directly modeling the treatment-covariate interaction. However, when the variance of the outcome is heterogeneous with respect to treatment and covariates, D-Learning does not leverage this structure. Stabilized Direct Learning (SD-Learning), proposed in this paper, utilizes potential heteroscedasticity in the error term through a residual reweighting which models the residual variance via flexible machine learning algorithms such as XGBoost and random forests. We also develop an internal cross-validation scheme which determines the best residual model amongst competing models. SD-Learning improves the efficiency of D-Learning estimates in binary and multi-arm treatment scenarios. The method is simple to implement and an easy way to improve existing algorithms within the D-Learning family, including original D-Learning, Angle-based D-Learning (AD-Learning), and Robust D-Learning (RD-Learning). We provide theoretical properties and justification of the optimality of SD-Learning. Head-to-head performance comparisons with D-Learning methods are provided through simulations, which demonstrate improvement in terms of average prediction error (APE), misclassification rate, and empirical value, along with data analysis of an AIDS randomized clinical trial.\\
\end{abstract}

%  Please place your key words in alphabetical order, separated
%  by semicolons, with the first letter of the first word capitalized,
%  and a period at the end of the list.
%

%  As usual, the \maketitle command creates the title and author/affiliations
%  display 

\maketitle

%  If you are using the referee option, a new page, numbered page 1, will
%  start after the summary and keywords.  The page numbers thus count the
%  number of pages of your manuscript in the preferred submission style.
%  Remember, ``Normally, regular papers exceeding 25 pages and Reader Reaction 
%  papers exceeding 12 pages in (the preferred style) will be returned to 
%  the authors without review. The page limit includes acknowledgements, 
%  references, and appendices, but not tables and figures. The page count does 
%  not include the title page and abstract. A maximum of six (6) tables or 
%  figures combined is often required.''

%  You may now place the substance of your manuscript here.  Please use
%  the \section, \subsection, etc commands as described in the user guide.
%  Please use \label and \ref commands to cross-reference sections, equations,
%  tables, figures, etc.
%
%  Please DO NOT attempt to reformat the style of equation numbering!
%  For that matter, please do not attempt to redefine anything!

\section{Introduction} \label{sec_intro}

Precision medicine is a framework at the intersection of statistics, machine learning, and causal inference, for leveraging patient heterogeneity to improve
patient outcomes. Individualized treatment rules (ITR), attempting to maximize expected treatment benefit across a population, may recommend personalized actions based on patient demographic information, clinical biomarkers, or genetic data. One of the primary goals of precision medicine has been formalized as decision support - the estimation of optimal and near-optimal regimes (\cite{kosorok_precision_2019}). In this vein, it is important to develop algorithms which work efficiently with the data at hand.

In recent years, an extensive literature has been developed in the area of estimating optimal ITRs. Traditionally, algorithms have fallen into one of two categories: \textit{model-based} vs. \textit{policy-search} approaches. Model-based approaches may also be considered ``regression-based" or ``indirect" in that they first model the conditional response of interest and then invert the relationship between patient covariates, treatment, and outcome to estimate an optimal rule (\cite{kosorok_adaptive_2016}). Primary examples of model-based approaches are Q-Learning (\cite{qian_performance_2011}) and A-Learning (\cite{murphy_optimal_2003, robins_optimal_2004}). Q-Learning approaches model the outcome conditional on covariates, whereas A-Learning approaches model regret functions or contrast functions between treatments (\cite{schulte_q-_2014}). Policy-search approaches, on the other hand, maximize value functions directly instead of modeling the conditional mean. For this reason, they are also known as ``value-search" or ``direct-search" (\cite{kosorok_adaptive_2016}). By sidestepping the modeling step and directly searching for an optimal rule among a class of policies, they may avoid model misspecification (\cite{xiao_robust_2019}). Many policy-search approaches have reframed the maximization of clinical outcomes as a weighted classification problem with the goal of minimizing weighted classification error, including the outcome weighted learning (OWL) family of methods (\cite{zhao_estimating_2012, zhou_residual_2017, zhang_multicategory_2020}) and various tree-based extensions (\cite*{cui_tree_2017}; \cite{kallus_balanced_2018, zhu_greedy_2017}).

A third category of algorithms to estimate ITRs also exists, consisting of methods which have attempted to combine the advantages of model-based and policy-search approaches. Often, these methods use the augmented inverse probability weighted estimator (AIPWE) within the classification framework. Such approaches are ``doubly robust" in the sense that they enjoy greater protection against model misspecification and increased efficiency when both models are correctly specified (\cite{zhao_efficient_2019, zhang_c-learning_2018, liu_augmented_2018}). 

\cite{tian_simple_2014} developed a ``modified-covariate" approach for ITR estimation, which falls into the third category but is unique in that it maintains the regression-based framework to model the treatment-covariate interaction effect directly, without having to specify a main effect model or conditional mean outcome function. Later, \cite{qi_d-learning_2018} coined the term ``Direct Learning" (D-Learning), which extended \cite{tian_simple_2014}'s idea to nonlinear decision rules and multi-arm treatment settings. Key extensions to D-Learning are Angle-based Direct Learning (AD-Learning) (\cite{qi_multi-armed_2020}), which improves D-Learning in the multi-arm treatment case, and Robust Direct Learning (RD-Learning) (\cite{meng_doubly_2021}),  which achieves a double robustness property.

We provide motivation for the ITR estimation approach of this paper by considering data from the ``AIDS Clinical Trial Group Study 175" (ACTG175), a randomized clinical trial (RCT) which compared the effectiveness of four treatments in increasing CD4 cell counts in HIV-1 patients (\cite{hammer_trial_1996}). Previous studies have suggested that the response of CD4 cell count from this data has skewed, heteroscedastic errors (\cite{zhang_robust_2021}) which may be dependent on patient age (\cite{xiao_robust_2019}). In such a situation, when the variance of the clinical outcome is a function of the covariates (or treatment), the D-Learning family of estimators remains consistent for the optimal ITR, but gives each observation equal weight by default in model training. A reweighting approach which utilizes this error structure to prioritize observations with smaller expected outcome variance is beneficial because it can attain greater efficiency when estimating an ITR.

In this article, we propose Stabilized D-Learning (SD-Learning), a method to increase the efficiency of D-Learning estimates in situations where the variance of the error term is non-homogeneous and a function of the treatment and covariates. SD-Learning leverages this heteroscedasticity through a residual reweighting framework where residual variance is modeled through flexible machine learning methods. We also develop an internal cross-validation scheme allowing for the selection of an optimal model amongst methods such as XGBoost (\cite{chen_xgboost_2016}) and random forests (\cite{breiman_random_2001}). The extension of SD-Learning to scenarios with $K \geq 3$ treatment arms is also included, and estimation fits within a least squares framework.

SD-Learning is simple to implement and can be stacked on top of existing D-Learning approaches to improve efficiency in a range of situations. We show that SD-Learning parameter estimates are consistent, achieve asymptotic normality in binary and multi-arm treatment scenarios in the presence of heterogeneous error, and have greater efficiency than D-Learning estimates. Additionally, we establish value function convergence bounds. In our simulations, we add a residual reweighting step to D-Learning, AD-Learning, and RD-Learning to show that SD-Learning can improve parameter estimation across a variety of scenarios with different error structures.

The rest of this paper is organized as follows: In Section \ref{sec_sdlearn}, we introduce the methodology of SD-Learning. Specifically, Section \ref{subsec_dlearn} reviews recent developments in D-Learning and Section \ref{subsec_sdlearn} introduces the mathematical motivation behind SD-Learning and outlines the reweighting solution. Section \ref{subsec_sdlearn_multi} extends the reweighting solution to scenarios with multi-arm treatments. In Section \ref{subsec_sdlearn_residual}, the residual model fitting step of the method is described in greater detail, and a stepwise implementation of the method is delineated. Section \ref{subsec_sdlearn_observational} extends the method to ITR estimation settings with observational data. In Section \ref{sec_theory}, theoretical results for SD-Learning including consistency, asymptotic normality, asymptotic efficiency, and value bounds are established for binary and multi-arm settings. Head-to-head simulations comparing SD-Learning to D-Learning, AD-Learning, and RD-Learning based on average prediction error (APE), misclassification rate, and empirical value are provided in Section \ref{sec_simulations}, and value comparisons from analysis of the ACTG175 RCT data are made in Section \ref{sec_data}. Concluding discussions and areas for future work are presented in Section \ref{sec_discussion}.

\section{Stabilized Direct Learning (SD-Learning)} \label{sec_sdlearn}

Although SD-Learning can be used with observational data, for simplicity, we first consider an RCT setting to demonstrate the methodology. For $n$ patients, we observe independent realizations of the random triplet $(X, A, R)$. Patient covariates are represented by the $p$-dimensional vector, $X \in \mathcal{X} \subset \mathbb{R}^{p}$, which includes an intercept. We start with the binary treatment scenario, $A \in \mathcal{A} = \{ -1, 1 \}$. A patient's clinical outcome is represented by $R \in \mathbb{R}$, and it is assumed, without loss of generality, that larger $R$ corresponds to a better outcome. The probability of receiving treatment $a$, given covariates $x$, is represented by $\pi(a, x) = \Pr(A=a|X=x)$. An ITR, $d(X) \colon \mathcal{X} \mapsto \mathcal{A}$, is a mapping from the space of covariates to the the space of treatments. We use $\mathbbm{1}(\cdot)$ to represent the indicator function. For a matrix $Z$, let $Z^{\top}$ denote its transpose. $\mathbb{P}_n \left( \cdot \right)$ represents the empirical average such that $\mathbb{P}_n(X) = n^{-1} \sum_{i=1}^{n} x_i$, where $x_1,...,x_n$ are realizations of the random variable, $X$. Let $\mbox{Vec}(Z)$ represent vectorization of $Z$, which stacks the columns (e.g. for $Z = \big(\begin{smallmatrix}
a & b \\ c & d \end{smallmatrix}\big)$, $\mbox{Vec}(Z) = [a\ c\ b\ d]^{\top}$).

Let $Y^*(-1)$ and $Y^*(1)$ represent the potential outcome that would have been observed had a patient received treatment $-1$ or $1$, respectively. We make the usual assumptions for this context from the potential outcomes framework of \cite{rubin_estimating_1974}: (i) Stable unit treatment value assumption (SUTVA): $Y = Y^*(A)$, (ii) No unmeasured confounding (conditional exchangeability): $A \perp \left\{ Y^*(-1), Y^*(1) \right\} \mid X$, and (iii) Positivity: $\pi(A,X) > c > 0, \forall\ A \in \mathcal{A}, X \in \mathcal{X}$. Prior to outlining the proposed SD-Learning methodology, we review key findings from the D-Learning family of methods.

\subsection{D-Learning Background} \label{subsec_dlearn}

It is known from \cite{qian_performance_2011} that the expected response under an ITR, $d$, can be represented by the value function:
\begin{equation*}
V(d) = E\left\{ R \mid A=d(X) \right\} = E \left[ \frac{R \cdot \mathbbm{1}\{A=d(X)\}}{\pi(A,X)} \right],
\end{equation*}
and we define an optimal ITR, $d^{opt}$, as the decision rule that maximizes the expected average response: $d^{opt}(\cdot) = \underset{d \in \mathcal{D}}{\mathrm{argmax}}\ V(d)$, where $\mathcal{D}$ is a prespecified class of decision rules. 

\subsubsection{D-Learning} \label{subsubsec_dlearn}

In the two-arm setting, assume that the outcome can be expressed by:
\begin{equation} \label{binary_working_model}
    R = m(X) + \delta(X)A + \eta,
\end{equation}
where $m(X)$ represents the main effect, $\delta(X)$ represents the treatment interaction effect, and $\eta$ is a mean-zero random error term. Note the following:

\begin{align}
\begin{split}
d^{opt}(X) &= \mbox{sign} \left\{ E(R | X, A=1) - E(R | X, A=-1) \right\} \\ &\coloneqq \mbox{sign} \left\{ f^{opt}(X) \right\}, 
\end{split} \label{opt_decision_rule} \\
\begin{split}
    f^{opt}(X) &= E \left\{ \frac{RA}{\pi(A,X)} \bigg| X \right\} = 2\delta(X). \label{opt_decision_function}
\end{split}
\end{align}

\cite{tian_simple_2014} made the connection between the optimal ITR in (\ref{opt_decision_rule}) and formulation of the optimal decision function in (\ref{opt_decision_function}), which forms the basis of D-Learning, as $f^{opt}(X)$ can now be directly learned through a regression method of choice. Lemma 1 of \cite{qi_d-learning_2018} shows that an estimation framework for $f^{opt}(X)$ in (\ref{opt_decision_function}) is:
\begin{equation} \label{dlearning}
 f^{opt}(X) \in \underset{f}{\mathrm{argmin}}\ E \left[ \frac{\left\{ 2RA - f(X) \right\}^2}{\pi(A,X)} \right],   
\end{equation}
and considering the class of linear decision functions, $\mathcal{F} = \{ f(X) = X^{\top}\beta : \beta \in \mathbb{R}^{p} \}$, the estimation problem can be solved with ordinary least squares (OLS) with or without regularization.

\subsubsection{AD-Learning} \label{subsubsec_adlearn}

\cite{qi_d-learning_2018} proposed pairwise D-Learning for the case where $A \in \{ 1, 2, ..., K \}$. This was improved with AD-Learning (\cite{qi_multi-armed_2020}), which uses the angle-based approach of \cite{zhang_multicategory_2014} to project treatment $A$ into $K$ simplex vertices defined in $\mathbb{R}^{K-1}$. Let treatment $A$ be represented by the vector $u_A \in \mathbb{R}^{K-1}$:
\begin{equation} \label{angle_based_vectors}
u_A = \left\{
  \begin{array}{lr} 
      \frac{1}{\sqrt{K-1}} \mathbf{1}_{K-1}, & A=1 \\
      \sqrt{ \frac{K}{K-1}} e_{A-1} - \frac{1+\sqrt{K}}{\sqrt{(K-1)^3}} \mathbf{1}_{K-1}, & 2 \leq A \leq K.
  \end{array}
\right.
\end{equation}
Here, $e_{i}$ is a $(K-1)$-dimensional vector of zeroes with $1$ in the $i^{th}$ location. Let the random vector $U$ be such that $U \mid (X, A) \overset{a.s.}{=} u_A$. The working model is:
\begin{equation*}
    R = \mu(X) + \sum_{k=1}^K \delta_k(X)\mathbbm{1}(A=k) + \eta,
\end{equation*}
where $\mu(X)$ is the main effect, $\delta_k(X)$ is the interaction effect between the $k^{th}$ treatment and covariates, and $\eta$ is the mean-zero random error. The optimal ITR can then be expressed as:
\begin{equation} \label{ad_learning_itr}
\begin{split}
    d^{opt}(X) &= \underset{k \in \{1,...,K\}}{\mathrm{argmax}}\ E(R | X=x, A = k) \\
    &= \underset{k \in \{ 1,...,K\}}{\mathrm{argmax}}\ u_k^{\top} E \left\{ \frac{RU}{\pi(A, X)}  \bigg| X \right\} \\
    &= \underset{k \in \{ 1,...,K\}}{\mathrm{argmax}}\ u_k^{\top} \sum_{k=1}^K \delta_k(X)u_k \\
    &\coloneqq \underset{k \in \{ 1,...,K\}}{\mathrm{argmax}}\ u_k^{\top} f^{opt}(X) \\
    &= \underset{k \in \{1,...,K\}}{\mathrm{argmax}}\ \delta_k(X),
\end{split}
\end{equation}
where $f^{opt}(X) \colon \mathbb{R}^{p+1} \mapsto \mathbb{R}^{K-1}$. As shown in Lemma 1 of \cite{qi_multi-armed_2020}, this leads to an estimation problem for independent responses of the form:
\begin{equation*}
f^{opt}(X) \in \underset{f \in \mathbb{R}^{K-1}}{\mathrm{argmin}} E \left[ \frac{\left\{KRU-f(X)\right\}^{\top} \left\{KRU-f(X)\right\}}{\pi(A,X)} \right].
\end{equation*}
Importantly, this is equivalent to the following working model and estimation framework:
\begin{gather}
\frac{K}{K-1}R = U^{\top}f(X) + \epsilon, \label{multi_working_model} \\ 
f^{opt}(X) \in \underset{f \in \mathbb{R}^{K-1}}{\mathrm{argmin}}\ E \left[ \frac{1}{\pi(A,X)} \left\{ \frac{K}{K-1}R - U^{\top}f(X) \right\}^2 \right]. \nonumber
\end{gather}

\subsubsection{RD-Learning} \label{subsubsec_rdlearn}

\cite{meng_doubly_2021} develop RD-Learning, which replaces the outcome $r_i$ in D-Learning with the residual $r_i - \hat{m}(x_i)$, where $\hat{m}(X)$ is an estimator for the main effect, $m(X)$ (similarly to \cite{zhou_residual_2017}). This reduces the variance and leads to doubly robust estimation of the treatment effect in the sense that consistency is guaranteed if either the main effect model or propensity score model is correctly specified.

\subsection{SD-Learning} \label{subsec_sdlearn}

For the binary treatment RCT setting, $\pi(A,X)$ is known, and assuming (\ref{binary_working_model}), the D-Learning estimation problem in (\ref{dlearning}) induces the following working model:
\begin{equation*}
2RA = f(X) + \epsilon.
\end{equation*}
Assume that $E(\epsilon|A,X) = 0$ and $var(\epsilon|A,X) = \sigma_0^2(A,X)$. Note that this error term is very general; it can be an arbitrary function of the treatment and covariates. In this case, the D-Learning estimator of the treatment effect is consistent, but due to the potential heteroscedasticity, it may lack efficiency as it gives each observation equal weight. Considering decision functions in $\mathcal{F}$, we propose a modified D-Learning objective function based on reweighting to gain efficiency:
\begin{equation} \label{sd_objective_binary}
\hat{\beta}_n^S = \underset{\beta \in \mathbb{R}^{p}}{\mathrm{argmin}}\ \mathbb{P}_n \left\{ \frac{(2RA - X^{\top}\beta)^2}{w(A,X) \pi(A,X)} \right\},
\end{equation}
where $w(A,X)$ is an arbitrary set of weights which need to be specified and/or estimated. The following assumptions establish the basic conditions needed to find optimal weights. For all $A \in \mathcal{A}$ and $X \in \mathcal{X}$ almost surely:

\begin{assumption} \label{full_rank}
$E(XX^{\top})$ is full rank and $E \left\| X \right\|^2 < \infty$.
\end{assumption}

\begin{assumption} \label{nonzero_wts}
$0 < c_1 \leq \sigma_0^2(A,X) \leq c_2 < \infty$.
\end{assumption}

Assumption \ref{full_rank} imposes a finite second moment restriction and assumes nonsingularity of the covariates. Assumption \ref{nonzero_wts} ensures that the true residual variance function is finite and nonzero (bounded above and below). 

\begin{proposition} \label{weights_prop_binary}
Under Assumptions \ref{full_rank} and \ref{nonzero_wts}, setting $w(A,X) = \frac{\sigma_0^2(A,X)}{\pi(A,X)}$ minimizes the estimator of the asymptotic variance of (\ref{sd_objective_binary}).
\end{proposition}

Proposition \ref{weights_prop_binary} offers a simple way to perform the reweighting. Let $\hat{\beta}_n^D$ be a consistent estimate of $\beta_0$, which can be obtained by fitting a traditional D-Learning model (\cite{qi_d-learning_2018}). Since $\epsilon = 2AR - X^{\top} \beta$, $\sigma_0^2(A,X)$ can be estimated by regressing $\left(2AR - X^{\top}\hat{\beta}_n^D\right)^2$ on $(A,X)$ through a parametric or nonparametric model. The resulting prediction function can be denoted as $\hat{\sigma}_n^2(A,X)$. This procedure breaks down into the following implementation steps:
\begin{enumerate}
    \item Obtain a D-Learning estimator:
    \begin{equation*}
        \hat{\beta}_n^D = \underset{\beta \in \mathbb{R}^{p}}{\mathrm{argmin}}\ \frac{1}{n} \sum_{i=1}^n \frac{\left(2 r_i a_i - x_i^{\top}\beta\right)^2}{\pi(a_i, x_i)}.
    \end{equation*}
    \item Regress the squared residuals from Step 1, $\left(2AR - X^{\top}\hat{\beta}_n^D \right)^2$, on the treatment and covariates, $(A,X)$, to obtain prediction function $\hat{\sigma}_n^2(A,X)$.
    \item Find $\hat{\beta}_n^S$ using:
    \begin{equation} \label{beta_sdlearning_binary}
    \begin{split}
    \hat{\beta}_n^S &= \underset{\beta \in \mathbb{R}^{p}}{\mathrm{argmin}}\ \frac{1}{n} \sum_{i=1}^n \frac{\pi \left(a_i,x_i\right) \left(2r_i a_i - x_i^{\top}\beta \right)^2}{\hat{\sigma}_n^2 \left( a_i, x_i \right) \pi \left(a_i,x_i\right)} \\
    &= \underset{\beta \in \mathbb{R}^{p}}{\mathrm{argmin}}\ \frac{1}{n} \sum_{i=1}^n \frac{\left(2r_i a_i - x_i^{\top}\beta\right)^2}{\hat{\sigma}_n^2 \left( a_i, x_i \right)}.
    \end{split}
    \end{equation}
\end{enumerate}

Thus, the SD-Learning estimator for the binary treatment case is formulated as a least squares problem, reweighted by the inverse of the estimated residual variance. A procedure for obtaining an improved estimate of the parameters has therefore been provided for binary D-Learning in the case of heteroscedasticity. The same reweighting framework can be used in the case of RD-Learning, where the only differences are that a model for the main effect, $m(X)$, must be estimated, and the augmented outcome becomes $R^* = R - \hat{m}(X)$.

\subsection{Extension of SD-Learning to Multiple Treatments} \label{subsec_sdlearn_multi}

Now, we expand the treatment space to $K$ treatments, indexed as $A \in \{ 1, 2, ..., K \}$. Let $u_A \in \mathbb{R}^{K-1}$ be defined as per (\ref{angle_based_vectors}). We use the working model in (\ref{multi_working_model}) under the same scenario as Section \ref{subsec_sdlearn}: $E(\epsilon|A,X) = 0$ and $var(\epsilon|A,X) = \sigma_0^2(A,X)$. The class of linear decision functions is defined as $\mathcal{F} = \{ f(X) = B^{\top}X : B \in \mathbb{R}^{p \times (K-1)} \}$.

Adding an arbitrary weight term, $w(A,X)$, in the denominator, similarly to the binary case, we propose the SD-Learning objective function as a modified version of AD-Learning:
% \begin{equation}
% \label{sd_objective_multi1}
% \hat{B}_n^S = \underset{B \in \mathbb{R}^{p \times (K-1)}}{\mathrm{argmin}} \mathbbm{P}_n \bigg\{ \frac{1}{w(A,X) \pi(A,X)} \times \left( \frac{K}{K-1}R - U^{\top}B^{\top}X \right)^2 \bigg\}.
% \end{equation}
\begin{multline}
\label{sd_objective_multi1}
\hat{B}_n^S = \underset{B \in \mathbb{R}^{p \times (K-1)}}{\mathrm{argmin}} \mathbbm{P}_n \bigg\{ \frac{1}{w(A,X) \pi(A,X)} \\ \times \left( \frac{K}{K-1}R - U^{\top}B^{\top}X \right)^2 \bigg\}.
\end{multline}
Again, $w(A,X)$ must be optimally chosen. We can reframe this objective function so that it is easier to optimize. Using the identity $\mbox{Vec}\left( ABC \right) = \left( C^{\top} \otimes A \right) \mbox{Vec}(B)$ and the fact that $U^{\top}B^{\top}X$ is a scalar,
\begin{equation*}
\begin{split}
U^{\top}B^{\top}X &= \mbox{Vec} \left( U^{\top}B^{\top}X \right) \\
&= \left( X^{\top} \otimes U^{\top} \right) \mbox{Vec}\left( B^{\top} \right) \\
    &= X_*^{\top} B_*,
\end{split}
\end{equation*}
where $X_* = \left(X^{\top} \otimes U^{\top} \right)^{\top}$, $B_* = \mbox{Vec} \left( B^{\top} \right)$, and $\otimes$ denotes the Kronecker product. This allows for an equivalent reformulation of the SD-Learning estimation problem:
% \begin{equation} \label{sd_objective_multi2}
% \hat{B}_n^S = \underset{B \in \mathbb{R}^{p \times (K-1)}}{\mathrm{argmin}} \mathbbm{P}_n \bigg\{ \frac{1}{w(A,X) \pi(A,X)} \times \left( \frac{K}{K-1}R - X_*^{\top} B_* \right)^2 \bigg\}.
% \end{equation}
\begin{multline} \label{sd_objective_multi2}
\hat{B}_n^S = \underset{B \in \mathbb{R}^{p \times (K-1)}}{\mathrm{argmin}} \mathbbm{P}_n \bigg\{ \frac{1}{w(A,X) \pi(A,X)} \\ \times \left( \frac{K}{K-1}R - X_*^{\top} B_* \right)^2 \bigg\}.
\end{multline}
Note that $B_*$ and $X_*$ in (\ref{sd_objective_multi2}) are vectors in $\mathbb{R}^{p(K-1)}$, unlike $B$ and $X$ in (\ref{sd_objective_multi1}), which are a matrix in $\mathbb{R}^{p \times (K-1)}$ and vector in $\mathbb{R}^{p}$, respectively. $w(A,X)$ can now be optimized in a fashion akin to the binary SD-Learning case:

\begin{proposition} \label{weights_prop_multi}
Under Assumption \ref{full_rank} for $X_*$ instead of $X$ and Assumption \ref{nonzero_wts}, setting $w(A,X) = \frac{\sigma_0^2(A,X)}{\pi(A,X)}$ minimizes the estimator of the asymptotic variance of (\ref{sd_objective_multi2}). 
\end{proposition}

Note that these are the same weights as found in Proposition \ref{weights_prop_binary} for the binary treatment case. Having found the optimal weights, $w(A,X)$, we switch back to non-vectorized notation (using $U^{\top}B^{\top}X$ instead of the equivalent $X_*^{\top} B_*$). $\sigma_0^2(A,X)$ can be estimated by regressing $\left\{\frac{K}{K-1}R - u^{\top} \big(\hat{B}_n^{AD}\big)^{\top} X  \right\}^2$ on $(A,X)$ through a parametric or nonparametric model, with the estimate denoted by  $\hat{\sigma}_n^2(A,X)$. Let $\hat{B}_n^{AD}$ represent a consistent estimate of $B_0$, obtained via an AD-Learning model (\cite{qi_multi-armed_2020}). The implementation of this procedure is as follows:

\begin{enumerate}
    \item Obtain an AD-Learning estimator:
    \begin{equation*}
        \hat{B}_n^{AD} = \underset{B \in \mathbb{R}^{p \times (K-1)}}{\mathrm{argmin}}\ \frac{1}{n} \sum_{i=1}^n \frac{1}{\pi(a_i, x_i)} \left(\frac{K}{K-1} r_i - u_{a_i}^{\top} B^{\top} x_i\right)^2.
    \end{equation*}
    \item Regress the squared residuals from Step 1, $\left\{ \frac{K}{K-1}R - u^{\top} \left(\hat{B}_n^{AD}\right)^{\top} X  \right\}^2$, on $(A,X)$, to obtain prediction function $\hat{\sigma}_n^2(A,X)$.
    \item Find $\hat{B}_n^S$ using:
    \begin{equation} \label{beta_sdlearning_multi}
    \hat{B}_n^S = \underset{B \in \mathbb{R}^{p \times (K-1)}}{\mathrm{argmin}} \frac{1}{n} \sum_{i=1}^{n} \frac{1}{\hat{\sigma}_n^2(a_i, x_i)} \left(\frac{K}{K-1} r_i - u_{a_i}^{\top} B^{\top} x_i \right)^2.
    \end{equation}
\end{enumerate}

Thus, SD-Learning in the multi-arm scenario remains a least squares problem, and under non-homogeneous error structures, provides an increased-efficiency estimation approach through the angle-based framework (refer to Theoretical Results in Section \ref{sec_theory}). 

As per covariate dimensionality and sparsity assumptions, the OLS steps of the implementation (Steps 1 and 3) in the binary or multi-arm case can be replaced with LASSO, Ridge, Elastic Net, or other regularized least squares techniques. Detailed proofs of Propositions \ref{weights_prop_binary} and \ref{weights_prop_multi} can be found in Web Appendix A of the Supporting Information.

\subsection{Residual Model Fitting} \label{subsec_sdlearn_residual}

\begin{table*} %[ht!] (on non-biom version)
    \caption{High-level overview of the SD- and SAD-Learning algorithms.}
    \label{algorithm_table}
    %\small
    \centering
    \begin{tabular}{p{0.25in}p{6.25in}}
    \hline \hline
        (1) & \textbf{Initial Estimate:} Obtain a consistent estimate of the parameters of the decision function through D-Learning in the binary treatment case or AD-Learning in the multi-arm treatment case (unweighted). \\
        (2) & \textbf{Hyperparameter Tuning:} Obtain squared residuals from Step (1) and find optimal LASSO, random forest, XGBoost, and/or SuperLearner parameters for predicting squared residuals from covariates. \\
        (3) & \textbf{Internal CV:} Using K-fold cross-validation, compare average test set MSEs for methods in Step (2) and pick the method with the lowest error. \\
        (4) & \textbf{Obtaining Weights:} Using optimal method from Step (3), obtain the predicted squared residual for each of $n$ observations. \\
        (5) & \textbf{Reweighted Estimate:} Using predictions from Step (4) as weights, obtain stabilized parameter estimates through SD-Learning as in (\ref{beta_sdlearning_binary}) or SAD-Learning as in (\ref{beta_sdlearning_multi}). \\
    \hline \hline \\
    \end{tabular}
\end{table*}

The residual modeling step in the implementation of SD-Learning is important, with various parametric and nonparametric options. We propose LASSO, random forest, and/or tree-based XGBoost for residual modeling in order to include a diversity of approaches through parametric assumptions, bagging, and/or boosting, respectively. Additionally, all three methods were chosen for their speed, LASSO and XGBoost for their ability to handle sparsity (\cite{zhang_sparsity_2008}, \cite{fauzan_accuracy_2018}, \cite{chen_xgboost_2016}), and in the case of random forests and XGBoost, flexibility. They also have a relatively low number of hyperparameters to tune, making their implementation easier. A SuperLearner algorithm can also be used, which combines candidate parametric and nonparametric methods to find an optimal combination which minimizes cross-validated risk. The SuperLearner has been shown to perform asymptotically as well as or better than any of the constituent candidate learners (\cite{van_der_laan_super_2007}).

Instead of picking one residual modeling framework, we propose comparing multiple models using an internal cross-validation step. After residuals from the initial D-Learning fit are obtained, they are squared and regressed against the covariates, and hyperparameters for each model are tuned. Then, using mean squared prediction error of the squared residuals as the evaluation metric, the models can be compared using K-fold cross-validation. We describe the algorithm with the high-level framework given in Table \ref{algorithm_table}.

\subsection{Extension to Observational Data} \label{subsec_sdlearn_observational}

It was assumed in the development of the methodology that the data stem from an RCT setting. In practice, however, it is often the case that drug efficacy is evaluated retrospectively through observational data. In this case, treatment assignment probabilities are not known. If the conditional exchangeability assumption holds, the SD-Learning methodology remains intact, but $\pi(A,X)$ must be estimated given the observed covariates. As discussed in \citet*{chen_personalized_2016}, the estimate $\hat{\pi}(A,X)$ could be obtained by a parametric model such as logistic regression (multinomial regression in the multi-arm case), or a nonparametric method such as boosting, random forests, or support vector regression (SVR). 

\section{Theoretical Results} \label{sec_theory}

In this section, we establish the theoretical properties of SD-Learning for settings with fixed dimension, $p$. We establish consistency and asymptotic normality of the SD-Learning estimator, along with bounds for the empirical value function. Detailed proofs for all theorems and lemmas can be found in Web Appendix B of the Supporting Information. We state two additional assumptions below, and will delineate which assumptions are needed for each theorem.

\begin{assumption} \label{unif_consistency}
$\hat{\sigma}_n^2(A,X)$ is uniformly consistent for $\sigma_0^2(A,X)$. That is, $||\hat{\sigma}_n^2(A,X) - \sigma_0^2(A,X)||_{\infty, \left( \mathcal{A}, \mathcal{X} \right)} \overset{P}{\rightarrow} 0$, where $|| \cdot ||_{\infty, \left( \mathcal{A}, \mathcal{X} \right)}$ represents the uniform norm over $(\mathcal{A}, \mathcal{X})$, and $(\mathcal{A}, \mathcal{X})$ is in a bounded set. In the case of observational data, we also require $||\hat{\pi}_n(A,X) - \pi_0(A,X)||_{\infty, \left( \mathcal{A}, \mathcal{X} \right)} \overset{P}{\rightarrow} 0$.
\end{assumption}

\begin{assumption} \label{asymp_normality_conditions}
$\forall\ \gamma > 0, \exists\ \mathcal{G}$ which is P-Donsker such that $\Pr \left\{ \frac{X \epsilon}{\hat{\sigma}_n^2(A, X)} \in \mathcal{G} \right\} > 1 - \gamma,\ \forall\ n$-large.
\end{assumption}

In Lemma \ref{lemma}, below, we propose two estimation methods for $\hat{\sigma}_n^2(A,X)$ and provide justification that they satisfy Assumptions \ref{unif_consistency} and \ref{asymp_normality_conditions}. Then, in Theorem \ref{consistency}, we establish consistency of the SD-Learning estimator in the binary treatment setting, and with consistency established, asymptotic normality of the estimator is shown in Theorem \ref{asymp_normality}. 

\begin{lemma} \label{lemma}
Estimating $\sigma_0^2(A,X)$ with (1) Linear regression with arbitrary features and (2) Random forests satisfies Assumptions \ref{unif_consistency} and \ref{asymp_normality_conditions}.
\end{lemma}

\begin{theorem} \label{consistency}
If Assumptions \ref{full_rank}-\ref{unif_consistency} are met, $\hat{\beta}_n^{S} \overset{P}{\rightarrow} \beta_0$.
\end{theorem}

\begin{theorem} \label{asymp_normality}
Let $Y^* = 2AY$, $E(Y^*|X) = X\beta_0$, and $\hat{\beta}_n^S$ be a consistent estimator of $\beta_0$. Denote $U_0 = E \left\{ \dfrac{XX^{\top}}{\sigma_0^2 \left(A, X \right)} \right\}$. If Assumption \ref{asymp_normality_conditions} is additionally met, then $\sqrt{n} \big( 
\hat{\beta}_n^S - \beta_0 \big)$ is asymptotically normal with variance $U_0^{-1}$.
\end{theorem}

$\hat{\beta}_n^S$ achieves the lower bound of the asymptotic variance shown in Proposition \ref{weights_prop_binary}, and is thus the optimal estimator for $\beta_0$ among the weighted choices of (\ref{sd_objective_binary}). Thus we have achieved consistency, convergence, and asymptotic efficiency.

Since Section \ref{subsec_sdlearn_multi} unifies the framework between SD-Learning in binary vs. multi-arm treatment scenarios, the extension of Theorems \ref{consistency} and \ref{asymp_normality} to multi-arm treatment are natural. This development is outlined in Theorem \ref{multi_theory}:

\begin{theorem} \label{multi_theory}
Let $X_* = \left(X^{\top} \otimes U^{\top} \right)^{\top}$, $B_* = \mbox{Vec} \left( B^{\top} \right)$, $Y^* = \frac{K}{K-1}R$ where $A \in \left\{ 1, 2, ..., K \right\}$, and $E\left( Y^* | X_* \right) = X_*^{\top} B_*$. Denote $U_0 = E \left\{ \dfrac{X_* X_*^{\top}}{\sigma_0^2 \left(A, X \right)} \right\}$. Under Assumption \ref{full_rank} for $X_*$ instead of $X$, Assumption \ref{nonzero_wts}, and Assumption \ref{unif_consistency}, $\hat{B}_*^S \overset{P}{\rightarrow} B_*$. Moreover, if Assumption \ref{asymp_normality_conditions} is additionally met, $\sqrt{n} \left( \hat{B}_*^S - B_* \right)$ is asymptotically normally distributed with variance $U_0^{-1}$.
\end{theorem}

Thus, all results established for the binary treatment case extend to the multi-arm setting. Combining the asymptotic normality result of Theorem \ref{multi_theory} with Theorem 1 of \cite{qi_multi-armed_2020} which shows that
\begin{equation*}
    V(d^{opt}) - V(\hat{d}_n) \leq \dfrac{2K(K-1)}{1-C(K)} \left( E \left\| f^{opt} - \hat{f}_n \right\|_2^2 \right)^{1/2},
\end{equation*}
 where $C(K)$ is a constant only depending on $K$, $\sqrt{n}$-convergence of $V(\hat{d}_n)$ to $V(d^{opt})$ is established.
 
\section{Numerical Results: Simulation Studies} \label{sec_simulations}

We perform head-to-head comparisons between SD-Learning and the D-Learning family of methods: D-Learning, AD-Learning, and RD-Learning. In all simulations, LASSO is used to obtain estimates of the decision function parameters (Steps 1 and 5 of Table \ref{algorithm_table}), and we use internal cross-validation to pick between LASSO, random forest, and XGBoost at the intermediate residual modeling steps (Steps 2 and 3 of Table \ref{algorithm_table}). Methods are evaluated based on three criteria:
\begin{enumerate}
    \item Average Prediction Error (APE): Coefficient accuracy determined by mean squared error of true vs. predicted decision functions. For binary simulation settings, $APE = n^{-1} \sum_{i=1}^n \big( x_i^{\top} \beta_0 - x_i^{\top} \hat{\beta}_n \big)^2$; for multi-arm settings,  $APE = n^{-1} \sum_{i=1}^n \big\{ f^{opt}(x_i) - \hat{f}(x_i) \big\}^2 = n^{-1} \sum_{i=1}^n \big\{ \sum_{k=1}^K \delta_k(x_i)u_k - \sum_{k=1}^K \hat{\delta}_k(x_i)u_k \big\}^2$.
    \item Misclassification Rate: \% incorrect treatment assignment.
    \item Empirical Value: $\hat{V}(d) = \dfrac{\mathbb{P}_n \left[ R \cdot \mathbbm{1}\{ A=d(X) \} / \pi(A,X) \right]}{\mathbb{P}_n \left[ \mathbbm{1}\{ A=d(X)\} / \pi(A,X) \right] }$.
\end{enumerate}
Better performance corresponds to lower APE, lower misclassification rate, and higher empirical value. In all simulations, performance based on these criteria is determined on a test data set with 10000 observations. 100 replications of each simulation setting are performed.

We compare all binary methods with $p = \{ 30, 60, 120 \}$ and multi-arm methods with $p = \{ 20, 40, 60 \}$. Simulated observations are independent with continuous covariates generated according to a U[-1,1] distribution. To allow for heteroscedasticity, the outcome is generated according to the working model in (\ref{binary_working_model}) for binary treatments and (\ref{multi_working_model}) for multi-arm treatments, but with $\epsilon$ generated according to $\sigma_0(X) * Z$, where $Z \sim N(0, 1)$ and $\sigma_0(X) > 0$. Here, non-constant $\sigma_0(X)$ introduces heteroscedasticity.

\subsection{Binary Treatment Simulations} \label{subsec_simulations_binary}

We compare SD-Learning and D-Learning with four simulation settings where $n=200$:

\begin{enumerate}
    \item $m(X) = 1 + 2X_1 + X_2 + 0.5X_3$; 
    
    \noindent $\delta(X) = 0.5(0.9 - X_1) $; $\sigma_0^2(X) = 1$.
    \item $m(X) = 1 + 12X_1 + 6X_2 + 3X_3$; 
    
    \noindent $\delta(X) = 4X_1$; $\sigma_0^2(X) = 0.25 + (X_2 + 1)^2$.    
    \item $m(X) = 1 + 10X_1 + 10X_2 + 20X_3 + 5X_4$; 
    
    \noindent $\delta(X) = 4(0.3 - X_1 - X_2)$; $\sigma_0^2(X) = 1 + 4X_3^2$.
    \item $m(X) = 1 + 10X_1 + 6X_1^2 - 6X_2^2 + 10X_3$; 
    
    \noindent $\delta(X) = 2X_2^2 + 1.5X_3 + 3X_4$; $\sigma_0^2(X) = .25 + (.3 - X_1 - X_2)^2$.
\end{enumerate}
Scenario (1) is very similar to the third linear decision boundary scenario in \cite{qi_d-learning_2018}, and has homoscedastic error. In this case, an intermediate residual reweighting step is not necessary since the optimal weights are $w(X) = 1$ by design, which is the default for D-Learning. Scenarios (2)-(4) introduce heteroscedasticity through the error term. Scenarios (2) and (3) meet linear decision function assumptions, and in both cases, the interaction effect has variables in common with the main effect but not with the error function. Finally, (4) is a nonlinear decision boundary scenario where SD-Learning is misspecified. This scenario will help test the robustness of SD-Learning.

\begin{figure*}
    \centering
    \includegraphics[width = 7.0in]{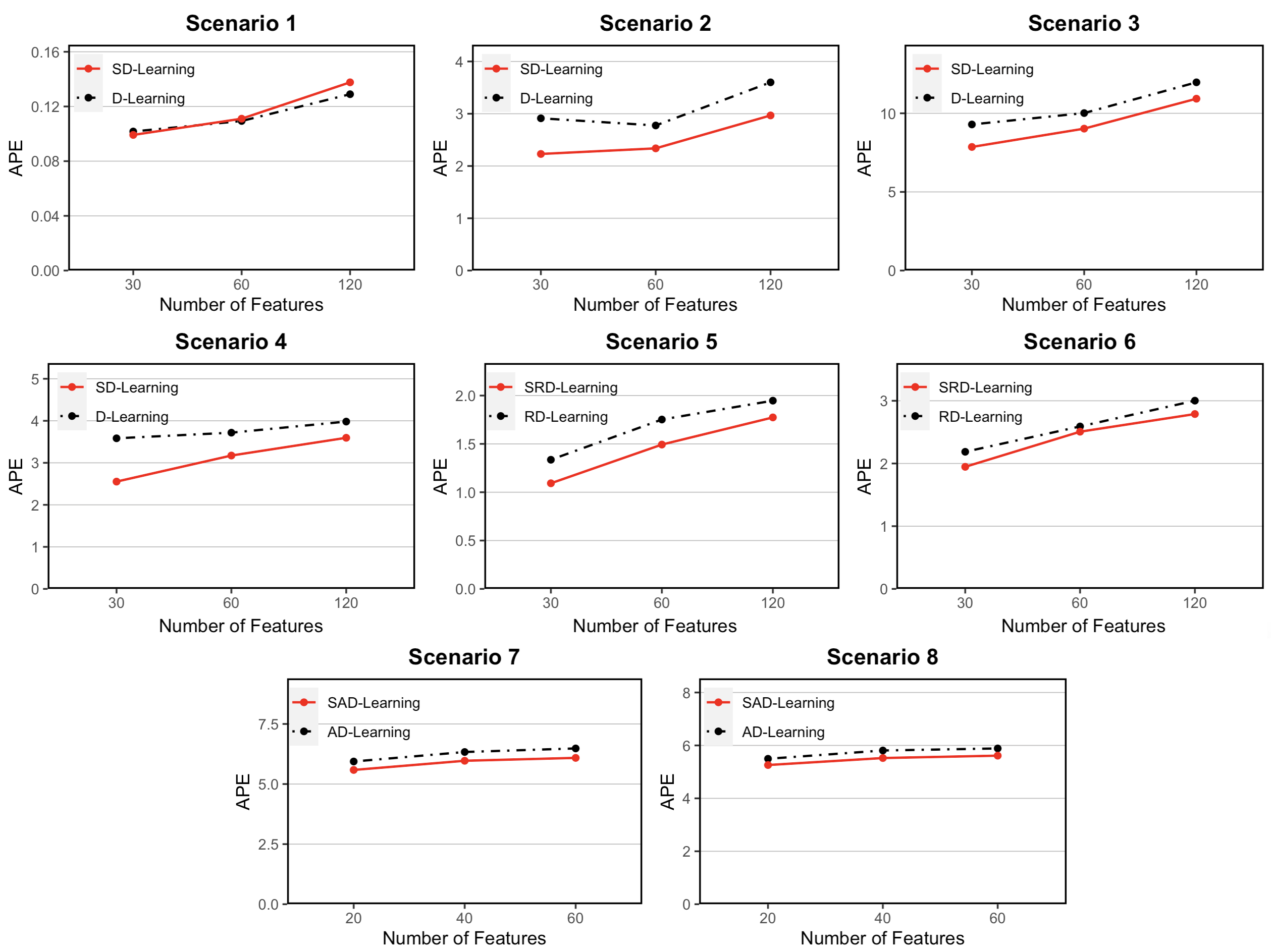}
    \caption{Average Prediction Error (APE) results of four binary simulation scenarios comparing D- to SD-Learning ($n=200$), two binary simulation scenarios comparing RD- to SRD-Learning ($n=100$), and two multi-arm simulation scenarios comparing AD- to SAD-Learning ($n=200$). In binary scenarios, $p$ varies from 30 to 120, and in multi-arm scenarios, $p$ varies from 20 to 60.}
    \label{sim_all_figure}
\end{figure*}

Figure \ref{sim_all_figure} shows the APE for all four scenarios, and Table \ref{sim_binary_d_table} contains misclassification rates and the estimated empirical value function on the test dataset. Since Scenario (1) has homoscedastic error, D-Learning performs ``optimally" in the sense of correctly specified weights; hence, stabilizing the D-Learning estimates is technically not needed. However, SD-Learning still performs similarly to D-Learning, with comparable APE and only slightly lower misclassification rate and empirical value. For scenarios (2) and (3), SD-Learning outperforms D-Learning because of the heterogeneous error structure. SD-Learning prioritizes observations with smaller expected outcome variance, and therefore estimates parameters more efficiently, shown by the lower APEs in Figure \ref{sim_all_figure}. This results in better classification and therefore, higher empirical value. In Scenario (4), although the true decision boundary is nonlinear, SD-Learning's flexible modeling of heteroscedasticity gives it an advantage over D-Learning.

\begin{table*}%[ht!]
    \caption{Mean empirical value and misclassification rate, along with standard error of the mean (SEM), for four binary D- vs. SD-Learning simulations and two binary RD- vs. SRD-Learning simulations for 30, 60, and 120 covariates. The best-performing method for each category is bolded.}
    \label{sim_binary_d_table}
    \centering
    \small
    \begin{tabular}{c c c c c c c c c}
    \hline
     & \multicolumn{2}{c}{$p=30$} & & \multicolumn{2}{c}{$p=60$} & & \multicolumn{2}{c}{$p=120$} \\ \cmidrule{2-3} \cmidrule{5-6} \cmidrule{8-9}
      & Value & Misclass. & & Value & Misclass. & & Value & Misclass.\\
     \Hline
     \multicolumn{9}{c}{Scenario 1} \\
    \hline
    D-Learning & \textbf{1.43} (0.01) & \textbf{0.07} (0.01) & & \textbf{1.43} (0.01) & \textbf{0.07} (0.01) & & \textbf{1.44} (0.01) & \textbf{0.08} (0.01)\\
    SD-Learning & 1.42 (0.01) & 0.08 (0.01) & & 1.42 (0.01) & 0.09 (0.01) & & 1.43 (0.01) & 0.09 (0.01)\\
    \hline
    \multicolumn{9}{c}{Scenario 2} \\
    \hline
    D-Learning & 2.27 (0.06) & 0.22 (0.01) & & 2.35 (0.07) & 0.21 (0.02) & & 2.30 (0.06) & 0.25 (0.02) \\
    SD-Learning & \textbf{2.46} (0.05) & \textbf{0.17} (0.01) & & \textbf{2.48} (0.06) & \textbf{0.18} (0.02) & & \textbf{2.50} (0.06) & \textbf{0.21} (0.01)\\
    \hline
    \multicolumn{9}{c}{Scenario 3} \\
    \hline
    D-Learning & 2.64 (0.09) & 0.29 (0.01) & & 2.40 (0.09) & 0.32 (0.01) & & 2.16 (0.08) & 0.36 (0.01)\\
    SD-Learning & \textbf{2.82} (0.08) & \textbf{0.26} (0.01) & & \textbf{2.56} (0.09) & \textbf{0.30} (0.01) & & \textbf{2.21} (0.08) & \textbf{0.35} (0.01) \\
    \hline
    \multicolumn{9}{c}{Scenario 4} \\
    \hline
    D-Learning & 1.81 (0.05) & 0.31 (0.01) & & 1.91 (0.05) & 0.32 (0.01) & & 1.90 (0.04) & 0.33 (0.01) \\
    SD-Learning & \textbf{2.08} (0.03) & \textbf{0.24} (0.01) & & \textbf{2.02} (0.05) & \textbf{0.29} (0.01) & & \textbf{1.99} (0.04) & \textbf{0.31} (0.01)\\
    \hline
    \multicolumn{9}{c}{Scenario 5} \\
    \hline
    RD-Learning & 1.94 (0.03) & 0.20 (0.01) & & 1.91 (0.05) & 0.27 (0.01) & & 1.43 (0.05) & 0.30 (0.01)\\
    SRD-Learning & \textbf{1.98} (0.03) & \textbf{0.18} (0.01) & & \textbf{2.02} (0.04) & \textbf{0.24} (0.01) & & \textbf{1.55} (0.05) & \textbf{0.27} (0.01)\\
    \hline
    \multicolumn{9}{c}{Scenario 6} \\
    \hline
    RD-Learning & 6.16 (0.04) & 0.21 (0.01) & & 5.94 (0.04) & 0.23 (0.01) & & 5.65 (0.05) & 0.25 (0.02) \\
    SRD-Learning & \textbf{6.24} (0.04) & \textbf{0.19} (0.01) & & \textbf{5.97} (0.04) & 0.23 (0.01) & & \textbf{5.70} (0.05) & \textbf{0.24} (0.01)\\
    \hline \hline \\
    \end{tabular}
\end{table*}

We also perform simulations to show the advantage of stabilizing the estimates of RD-Learning. For these simulations, $n=100$. For the main effect modeling step, LASSO was used, but nonparametric methods may also be used as per \cite{meng_doubly_2021}. The simulation settings for SRD-Learning vs. RD-Learning are as follows:
\begin{enumerate}
\setcounter{enumi}{4}
    \item $m(X) = 1 + 10X_1 + 10X_2 + 20X_3 + 20X_5 + 10X_1X_2$; 
    
    \noindent $\delta(X) = 1.25X_3 + 2.5X_4$; $\sigma_0^2(X) = 1+0.1(1 + X_1^2)$.
    \item $m(X) = 1 + 5\cos^2(X_1) + 10X_1X_2 + 20X_2 + 30X_5$; 
    
    \noindent $\delta(X) = 3X_1 + 2X_2 + 2X_5^2$; $\sigma_0^2(X) = 0.5 + 0.5\left( 1 - 0.25X_6 \right)^3$.
\end{enumerate}

Both scenarios have heterogeneous error. Scenario (5) has a true linear decision function, whereas (6) has a nonlinear decision function along with a nonlinear \textit{cosine} term in the main effect. 

Figure \ref{sim_all_figure} shows the APE for both scenarios, and Table \ref{sim_binary_d_table} contains misclassification rates and the estimated empirical value function on the test dataset. Although RD-Learning is already robust in the sense that the main effect is removed before model fitting, the reweighing step of SRD-Learning adds efficiency in situations with heteroscedasticity - even in the presence of a misspecified decision function and nonlinearity in the main effect.

\subsection{Multi-Arm Treatment Simulations} \label{subsec_simulations_multi}

We compare SAD-Learning and AD-Learning with two multi-arm treatment scenarios under heteroscedasticity, setting $K=4$ and $n=200$ in both:

\begin{enumerate}
\setcounter{enumi}{6}
    \item $m(X) = 1 + 2X_1 + 2X_2$; 
    
    \noindent $\sigma_0^2(X) = 0.25 + 0.2(1.5 - X_2)^2$;
    
    \noindent $\delta(X) = 
    \begin{cases} 
      .75 + 1.5X_1 + 1.5X_2 + 1.5X_3 + 1.5X_4, & A = 1 \\
      .75 + 1.5X_1 - 1.5X_2 - 1.5X_3 + 1.5X_4, & A = 2 \\
      .75 + 1.5X_1 - 1.5X_2 + 1.5X_3 - 1.5X_4, & A = 3 \\
      .75 - 1.5X_1 + 1.5X_2 - 1.5X_3 + 1.5X_4, & A = 4.
   \end{cases}$ \newline
   \item $m(X) = 1 + X_5 + 3X_6 + 2X_1X_2$; 
   
   \noindent $\sigma_0^2(X) = 0.5 + 2X_2*\mathbbm{1}(X_2 > 0)$;
   
   \noindent $\delta(X) = 
    \begin{cases} 
      0.5 + 2X_1 + X_2 + X_3, & A = 1 \\
      1 + X_1 - X_2 - X_3, & A = 2 \\
      1.5 + 3X_1 - X_2 + X_3, & A = 3 \\
      1 - X_1 - X_2 + X_3, & A = 4.
   \end{cases}$
\end{enumerate}

Both Scenarios (7) and (8) meet linear assumptions for the treatment-covariate interaction effects. In Scenario (7), the heterogeneous error function is a quadratic function of the covariate $X_2$, but in Scenario (8) it is a spline function of $X_2$ where the variance is constant for $X_2\leq0$ and increases thereafter. Figure \ref{sim_all_figure} reports the APE for both scenarios, and Table \ref{sim_multi_ad_table} displays misclassification and value results. As observed in the binary cases, reweighting in the multi-arm scenario under heteroscedasticity also improves efficiency, resulting in lower APEs, lower misclassification rates, and higher empirical values. SAD-Learning improves the performance of AD-Learning when an error structure can be learned and utilized to obtain decision rules built by favoring observations which are more likely to represent signal than noise.
 
\begin{table*}%[ht!]
    \caption{Mean empirical value and misclassification rate, along with standard error of the mean (SEM), for two multi-arm scenarios comparing AD-Learning to SAD-Learning. All simulations are repeated with 20, 40, and 60 covariates. The best-performing method for each category is bolded.}
    \label{sim_multi_ad_table}
    \centering
    \small
    \begin{tabular}{c c c c c c c c c}
    \hline
     & \multicolumn{2}{c}{$p=20$} & & \multicolumn{2}{c}{$p=40$} & & \multicolumn{2}{c}{$p=60$} \\ \cmidrule{2-3} \cmidrule{5-6} \cmidrule{8-9}
      & Value & Misclass. & & Value & Misclass. & & Value & Misclass.\\
     \Hline
     \multicolumn{9}{c}{Scenario 7} \\
    \hline
    AD-Learning & 3.39 (0.02) & 0.34 (0.01) & & 3.23 (0.03) & 0.39 (0.01) & & 3.19 (0.03) & 0.42 (0.01)\\
    SAD-Learning & \textbf{3.49} (0.01) & \textbf{0.30} (0.01) & & \textbf{3.39} (0.02) & \textbf{0.33} (0.01) & & \textbf{3.39} (0.02) & \textbf{0.35} (0.01)\\
    \hline
    \multicolumn{9}{c}{Scenario 8} \\
    \hline
    AD-Learning & 2.82 (0.04) & 0.46 (0.01) & & 2.81 (0.03) & 0.47 (0.01) & & 2.79 (0.04) & 0.50 (0.01) \\
    SAD-Learning & \textbf{3.02} (0.02) & \textbf{0.38} (0.01) & & \textbf{3.02} (0.02) & \textbf{0.39} (0.01) & & \textbf{2.95} (0.03) & \textbf{0.43} (0.01)\\
    \hline \\
    \end{tabular}
\end{table*}

\begin{table*}%[ht!]
    \caption{Empirical value estimates for binary AIDS data scenarios comparing performance of D- and SD-Learning on each pairwise set of treatments (Z, ZD, ZZ, D) at varying sample sizes of training data ($n=100, 200, 400, 800$). At each sample size, results are averaged from 1000 replications, and corresponding standard error of the mean (SEM) is shown. The best-performing method at each level of sample size is bolded. When both methods converge upon recommending a single treatment (in over 99\% of patients across all replications), the treatment is specified instead of the (nearly identical) value estimates.}
    \label{aids_binary_table}
    \centering
    %\scalebox{0.8}{
    \begin{tabular}{c c c c c c c c c}
    \hline
     & & D-Learning & SD-Learning & & & & D-Learning & SD-Learning \\
    \Hline
    ZD vs. ZZ & $n=100$ & 49.12 (0.31) & \textbf{49.24} (0.30) & & Z vs. ZD & $n=100$ & 51.75 (0.21) & \textbf{52.18} (0.20)\\
              & $n=200$ & 51.76 (0.21) & \textbf{52.12} (0.20) & & & $n=200$ & 52.92 (0.15) & \textbf{53.43} (0.14)\\
              & $n=400$ & 52.74 (0.18) & \textbf{53.15} (0.17) & & & $n=400$ & 53.48 (0.16) & \textbf{53.90} (0.16)\\
              & $n=800$ & 53.16 (0.35) & \textbf{53.55} (0.36) & & & $n=800$ & 53.84 (0.35) & \textbf{54.30} (0.35)\\
              \hline
    ZD vs. D & $n=100$ & 48.67 (0.29) & \textbf{48.92} (0.28) & & Z vs. ZZ & $n=100$ & 15.62 (0.27) & 15.62 (0.27)\\
              & $n=200$ & 50.59 (0.23) & \textbf{51.02} (0.23) & & & $n=200$ & 17.96 (0.14) & \textbf{18.10} (0.15)\\
              & $n=400$ & 52.88 (0.19) & \textbf{53.28} (0.19) & & & $n=400$ & \multicolumn{2}{c}{\textit{ZZ for over 99\% of patients.}}\\
              & $n=800$ & 55.99 (0.33) & \textbf{56.42} (0.33) & & & $n=800$ & \multicolumn{2}{c}{\textit{ZZ for over 99\% of patients.}}\\
              \hline
    ZZ vs. D & $n=100$ & 23.57 (0.13) & \textbf{23.64} (0.13) & & Z vs. D & $n=100$ & \textbf{24.57} (0.19) & 24.55 (0.19)\\
              & $n=200$ & \textbf{23.89} (0.13) & 23.88 (0.14) & & & $n=200$ & \multicolumn{2}{c}{\textit{D for over 99\% of patients.}}\\
              & $n=400$ & 24.52 (0.15) & \textbf{24.57} (0.15) & & & $n=400$ & \multicolumn{2}{c}{\textit{D for over 99\% of patients.}}\\
              & $n=800$ & \textbf{25.55} (0.25) & 25.48 (0.25) & & & $n=800$ & \multicolumn{2}{c}{\textit{D for over 99\% of patients.}}\\
    \hline \hline \\
    \end{tabular}
    %}
\end{table*}

\begin{table*}%[ht!]
    \caption{Empirical value estimates for multi-arm AIDS data scenarios comparing the performance of AD- and SAD-Learning in selecting amongst four treatments simultaneously. Varying sample sizes of the training data were chosen to be $n=100$, $200$, $400$, $800$, and $1200$. At each sample size, results are averaged from 1000 replications, and the corresponding standard error of the mean (SEM) is shown. The best-performing method at each level of sample size is bolded.}
    \label{aids_multi_table}
    \centering
    \small
    \begin{tabular}{c c c}
    \hline
      & AD-Learning & SAD-Learning \\
     \Hline
    $n=100$ & 43.27 (0.43) & \textbf{43.98} (0.43)\\
    $n=200$ & 47.19 (0.37) & \textbf{48.29} (0.35) \\
    $n=400$ & 50.57 (0.25) & \textbf{51.74} (0.22)\\
    $n=800$ & 53.35 (0.18) & \textbf{54.25} (0.17)\\
    $n=1200$ & 54.27 (0.23) & \textbf{55.24} (0.23)\\
    \hline \\
    \end{tabular}
\end{table*}

\section{Data Analysis: AIDS Clinical Trial} \label{sec_data}

We apply SD-Learning approaches with a linear decision function to ACTG175 data, which may benefit from a reweighting approach, as motivated in Section \ref{sec_intro}. The ACTG175 dataset is publicly available via the R package \textit{speff2trial}. This double-blinded study evaluated monotherapy vs. combination therapy approaches to increasing CD4 cell counts in HIV-1-infected patients with initial cell counts of between 200 and 500 cells/mm$^3$ (\cite{hammer_trial_1996}). AIDS-defining illnesses have been shown to decrease as CD4 cell count increases (\cite{mocroft_incidence_2013}), so larger increases in CD4 cell count are preferable. 

Patients were randomly assigned to one of four daily regimens with equal probability:
\begin{enumerate}
    \item 600 mg zidovudine (Z)
    \item 600 mg zidovudine + 400 mg didanosine (ZD)
    \item 600 mg zidovudine + 2.25 mg zalcitabine (ZZ)
    \item 400 mg didanosine (D)
\end{enumerate}

The outcome we use for this analysis is the change in CD4 cell count from baseline to 20 weeks, as done in \cite{qi_d-learning_2018} and \cite{qi_multi-armed_2020}. 12 covariates are selected as per (\cite{fan_concordance-assisted_2017}); five continuous: weight (kg), age (years), Karnofsky score (0-100), baseline CD4 count (cells/mm$^3$), baseline CD8 count (cells/mm$^3$); and seven binary: hemophilia (1=yes, 2=no), homosexual activity (1=yes, 0=no), history of intravenous drug use (1=yes, 0=no), race (1=non-white, 0=white), gender (1=male, 0=female), antiretroviral history (1=experienced, 0=naive), and symptomatic status (1=symptomatic, 0=asymptomatic). For all comparisons, LASSO is used to obtain estimates of the decision function parameters. LASSO, random forest, and XGBoost are tuned and used for the residual modeling step of SD- and SAD-Learning, with the optimal method picked by internal cross-validation.

\subsection{Binary Scenario} \label{subsec_data_binary}

We compare the performance of D-Learning and its corresponding stabilized version, SD-Learning, for each pairwise set of treatments from the four choices. We randomly split the data into a training set of $n$ observations, using the rest of the observations as testing data. $n$ is selected to be 100, 200, 400, and 800. For generating empirical value estimates, $\hat{V}(d)$, we perform Monte Carlo Cross-Validation (repeated random subsampling) with 1000 iterations at each $n$. The corresponding binary treatment results are shown in Table \ref{aids_binary_table}.

In terms of empirical value, SD-Learning improves upon the performance of D-Learning in most scenarios, especially for comparisons involving treatment ZD. SD- and D-Learning perform approximately equally well for ZZ vs. D, with empirical values differing by less than $0.10$ in all cases. For pairwise comparisons of Z vs. ZZ and Z vs. D, both methods eventually converge upon recommending a single treatment to over $99\%$ of patients and therefore have very similar value estimates. Overall, SD-Learning either outperforms D-Learning (ZD vs. ZZ, ZD vs. D, and Z vs. ZD) or performs equally as well (ZZ vs. D, Z vs. ZZ, Z vs. D).

\subsection{Multi-Arm Treatment Scenario} \label{subsec_data_multi}

We now compare the performance of AD- and SAD-Learning at varying sample sizes while considering all four treatments simultaneously. We randomly split the data into training sets of $n = 100, 200, 400, 800$, and $1200$ observations, using the rest of the observations as testing data. The procedure is repeated 1000 times for each value of $n$. All results are shown in Table \ref{aids_multi_table}.

We can see that SAD-Learning has a distinct advantage over AD-Learning at all observed training data sample sizes. The stabilization step weighs patients differentially based on predicted squared residual from the initial AD-Learning step. This results in more efficiently estimated treatment rules and therefore higher values, even in scenarios with very low sample sizes.

\section{Discussion} \label{sec_discussion}

In this article, we propose SD-Learning, which boosts the efficiency of D-Learning in a wide range of scenarios with more general error functions, thus enhancing the utility of D-Learning on datasets which may be encountered in practice. The performance of SD-Learning relies on sufficiently modeling residuals from an initial D-Learning fit, which can be achieved through a variety of parametric or nonparametric methods. 

From another perspective, SD-Learning may be considered an extension of feasible weighted least squares (FWLS) (\cite{olive_wls_2017}) to the precision medicine setting where optimal ITR estimation with two or more treatments is of interest. Similarly to FWLS, SD-Learning is motivated by efficient estimation and controls on variance (\cite{miller_feasible_2018}). Our results suggest that SD-Learning pays a minor efficiency price under homogeneous error, but can offer substantial efficiency gains in the case of heterogeneous error. Additionally, SD-Learning parameter estimates are asymptotically normal when OLS is used for the estimation steps, allowing post-modeling inference even in the multi-arm treatment scenario. 

The implementation benefits of SD-, SRD-, and SAD-Learning lie in the fact that they are straightforward to use and can simply be stacked on top of D-, RD-, and AD-Learning which have been shown to perform well in a multitude of settings. Additionally, our methodology even in the multi-arm treatment setting (SAD-Learning) remains estimable with a least squares framework and closed-form solution, not requiring the use of optimization algorithms.

For practical use, we suggest considering a variety of nonparametric prediction algorithms for the residual modeling step in order to gain an understanding of the heteroscedasticity structure of the dataset at hand. Random forest appeared to work well under a wide variety of scenarios, with the number of trees as the most important tuning parameter. Although RCT settings were considered in this article, for observational data, we recommend using SRD-Learning, since \cite{meng_doubly_2021} showed that incorporation of the mean outcome model results in protection against incorrect specification of the propensity score model. 

Several future extensions of this work are possible. Theoretical results in Section \ref{sec_theory}, including Lemma \ref{lemma} for random forests, can be extended to the case where covariate dimension, $p$, increases with $n$. A natural methodological extension is to allow for SD-Learning to estimate nonlinear decision rules using Reproducing Kernel Hilbert Space (RKHS) techniques. Additionally, SD-Learning may be broadened to work for binary and survival outcomes under heteroscedasticity. As proposed in this paper, SD-Learning assumes independence (but not identically distributed errors) between observations. It would also be valuable to use SD-Learning in scenarios with correlation between observations. This would be akin to feasible generalized least squares (FGLS) (\cite{olive_wls_2017}) for ITR estimation.

%  The \backmatter command formats the subsequent headings so that they
%  are in the journal style.  Please keep this command in your document
%  in this position, right after the final section of the main part of 
%  the paper and right before the Acknowledgements, Supporting Information (Supplementary %  Materials),   and References sections. 

\backmatter

\section*{Acknowledgements}

The authors thank Marissa Ashner and Kyle Grosser for taking time to review the manuscript and providing very helpful feedback.

%\newpage
\bibliographystyle{biom} \bibliography{references_bibtex}

%\section*{Supporting Information}
%
%Web Appendices referenced in Sections \ref{sec_sdlearn} and \ref{sec_theory} are available with this paper at the Biometrics website on Wiley Online Library.

\appendix

%  To get the journal style of heading for an appendix, mimic the following.

%\section{}
%\subsection{Title of appendix}

%Put your short appendix here.  Remember, longer appendices are
%possible when presented as Supplementary Web Material.  Please 
%review and follow the journal policy for this material, available
%under Instructions for Authors at %\texttt{http://www.biometrics.tibs.org}.

\label{lastpage}



\section*{Supplementary material}

# Supporting Information for "Stabilized Direct Learning for Efficient Estimation of Individualized Treatment Rules"

**Kushal S. Shah[1],\*, Haoda Fu[2],\*\*, and Michael R. Kosorok[1],\*\*\***

[1]Department of Biostatistics, University of North Carolina at Chapel Hill, NC 27599, USA

[2]Eli Lilly and Company, Lilly Corporate Center, Indianapolis, IN 46285, USA

*email:* kushshah@live.unc.edu

\*\**email:* fu_haoda@lilly.com

\*\*\**email:* kosorok@unc.edu

Web Appendix A

This section includes proofs of Propositions 1 and 2.

[**Proof of Proposition 1**] Let $l_n(\beta) = \mathbb{P}_n \left\{ \dfrac{(2RA - X^\top \beta)^2}{w(A, X)\pi(A, X)} \right\}$, and define the first and second derivatives of $l_n$:

$$\frac{\partial l_n(\beta)}{\partial \beta} = -2\mathbb{P}_n \left\{ \frac{(2RA - X^\top \beta)X}{w(A, X)\pi(A, X)} \right\} \tag{1}$$

$$\frac{\partial^2 l_n(\beta)}{\partial \beta^2} = 2\mathbb{P}_n \left\{ \frac{XX^\top}{w(A, X)\pi(A, X)} \right\} \tag{2}$$

Assume that $\hat{\beta} \xrightarrow{P} \beta$. Using Taylor Series approximations:

$$0 = \frac{\partial l_n(\hat{\beta})}{\partial \beta} = \frac{\partial l_n(\beta)}{\partial \beta} + \left( \frac{\partial^2 l_n(\beta)}{\partial \beta^2} + o_p(1) \right) \left( \hat{\beta} - \beta \right). \tag{3}$$

From this, we have:

$$\begin{aligned}
\sqrt{n}\left( \hat{\beta} - \beta \right) &= -\sqrt{n} \left\{ \frac{\partial^2 l_n(\beta)}{\partial \beta^2} \right\}^{-1} \frac{\partial l_n(\beta)}{\partial \beta} + o_p(1) \\
&= \sqrt{n} \left\{ \mathbb{P}_n \frac{XX^\top}{w(A, X)\pi(A, X)} \right\}^{-1} \left\{ \mathbb{P}_n \frac{(2AR - X^\top\beta)X}{w(A, X)\pi(A, X)} \right\} + o_p(1) \\
&= E \left[ \frac{XX^\top}{w(A, X)\pi(A, X)} \right]^{-1} \left\{ \sqrt{n}\mathbb{P}_n \frac{\epsilon X}{w(A, X)\pi(A, X)} \right\} + o_p(1), \text{ and}
\end{aligned} \tag{4}$$

$$\begin{aligned}
& var\left\{ \sqrt{n}\left( \hat{\beta} - \beta \right) \right\} \\
&= E \left[ \frac{XX^\top}{w(A, X)\pi(A, X)} \right]^{-1} var \left[ \frac{\epsilon X}{w(A, X)\pi(A, X)} \right] E \left[ \frac{XX^\top}{w(A, X)\pi(A, X)} \right]^{-1} + o(1) \\
&= E \left[ \frac{XX^\top}{w(A, X)\pi(A, X)} \right]^{-1} E \left[ \frac{XX^\top E\left( \epsilon^2 \mid A, X \right)}{w(A, X)^2 \pi(A, X)^2} \right] E \left[ \frac{XX^\top}{w(A, X)\pi(A, X)} \right]^{-1} + o(1) \\
&= E \left[ \frac{XX^\top}{w(A, X)\pi(A, X)} \right]^{-1} E \left[ \frac{XX^\top \sigma_0^2(A, X)}{w(A, X)^2 \pi(A, X)^2} \right] E \left[ \frac{XX^\top}{w(A, X)\pi(A, X)} \right]^{-1} + o(1) \\
&= A^{-1} B A^{-1} + o(1),
\end{aligned} \tag{5}$$

since $\sigma_0^2(A, X) = var\left( \epsilon \mid A, X \right) = E\left( \epsilon^2 \mid A, X \right)$. For ease of notation, let $A = E \left[ \dfrac{XX^\top}{w(A, X)\pi(A, X)} \right]$ and $B = E \left[ \dfrac{XX^\top \sigma_0^2(A, X)}{w(A, X)^2 \pi(A, X)^2} \right]$. Letting $w_t(A, X) = w(A, X) + ts(A, X)$ be a perturbation of the weights, we take a functional derivative and show that no matter the choice of $s(A, X)$, the first derivative of the variance expression is zero when $w(A, X) = \dfrac{\sigma_0^2(A, X)}{\pi(A, X)}$ (e.g. the

variance is minimized). The variance term can be written as:

$$
\begin{aligned}
&var\big\{\sqrt{n}(\hat{\beta} - \beta)\big\} \\
&= E\left[\frac{XX^\top}{w_t(A,X)\pi(A,X)}\right]^{-1} E\left[\frac{XX^\top \sigma_0^2(A,X)}{w_t(A,X)^2\pi(A,X)^2}\right] E\left[\frac{XX^\top}{w_t(A,X)\pi(A,X)}\right]^{-1} + o(1) \\
&= A_t^{-1} B_t A_t^{-1} + o(1).
\end{aligned}
$$

$$(6)$$

The derivatives of $A_t^{-1}$ and $B_t$ are:

$$
\frac{\partial A_t^{-1}}{\partial t} = -A_t^{-1}\frac{\partial A_t}{\partial t}A_t^{-1} = -A_t^{-1} E\left\{\frac{-XX^\top s(A,X)}{w_t(A,X)^2\pi(A,X)}\right\} A_t^{-1} = A_t^{-1} C_t A_t^{-1}, \tag{7}
$$

$$
\frac{\partial B_t}{\partial t} = E\left\{\frac{-2XX^\top \sigma_0^2(A,X)s(A,X)}{w_t(A,X)^3\pi(A,X)^2}\right\}, \tag{8}
$$

where $C_t = E\left\{\frac{XX^\top s(A,X)}{w_t(X)^2\pi(A,X)}\right\}$. Having noted this, we can take the first derivative of the

variance expression:

$$
\begin{aligned}
&\frac{\partial}{\partial t}\left[var\left\{\sqrt{n}\left(\hat{\beta} - \beta\right)\right\}\right] \\
&= \frac{\partial A_t^{-1}}{\partial t}B_t A_t^{-1} + A_t^{-1}\frac{\partial B_t}{\partial t}A_t^{-1} + A_t^{-1}B_t\frac{\partial A_t^{-1}}{\partial t} \\
&= A_t^{-1} C_t A_t^{-1} B_t A_t^{-1} + A_t^{-1} E\left\{\frac{-2XX^\top \sigma_0^2(A,X)s(A,X)}{w_t(A,X)^3\pi(A,X)^2}\right\} A_t^{-1} + A_t^{-1}B_t A_t^{-1} C_t A_t^{-1}.
\end{aligned}
$$

$$(9)$$

Therefore:

$$
\begin{aligned}
&\frac{\partial}{\partial t}\left[var\left\{\sqrt{n}\left(\hat{\beta} - \beta\right)\right\}\right]\bigg|_{t=0} \\
&= A^{-1}CA^{-1}BA^{-1} + A^{-1} E\left\{\frac{-2XX^\top \sigma_0^2(A,X)s(A,X)}{w(A,X)^3\pi(A,X)^2}\right\} A^{-1} + A^{-1}BA^{-1}CA^{-1}.
\end{aligned}
$$

$$(10)$$

Now, letting $w(A,X) = \frac{\sigma_0^2(A,X)}{\pi(A,X)}$:

$$
E\left\{\frac{-2XX^\top \sigma_0^2(A,X)s(A,X)}{w(A,X)^3\pi(A,X)^2}\right\} = -2C, \tag{11}
$$

$$
A^{-1}B = E\left\{\frac{XX^\top}{\sigma_0^2(A,X)}\right\}^{-1} E\left\{\frac{XX^\top}{\sigma_0^2(A,X)}\right\} = 1. \tag{12}
$$

Therefore:

$$
\begin{aligned}
\frac{\partial var\left\{\sqrt{n}(\hat{\beta} - \beta)\right\}}{\partial t}\bigg|_{t=0} &= A^{-1}CA^{-1} - 2A^{-1}CA^{-1} + A^{-1}CA^{-1} \\
&= 0.
\end{aligned}
$$

$$(13)$$

We have shown that the sandwich variance is minimized at $w(A,X) = \dfrac{\sigma_0^2(A,X)}{\pi(A,X)}$. Thus, the proof is complete.

[**Proof of Proposition 2**] Let $l_n(B_*) = \mathbb{P}_n \left\{ \dfrac{1}{w(A,X)\pi(A,X)} \left( \dfrac{K}{K-1} R - X_*^\top B_* \right)^2 \right\}$, and define the first and second derivatives of $l_n$:

$$\frac{\partial l_n(B_*)}{\partial B_*} = -2\mathbb{P}_n \left\{ \frac{1}{w(A,X)\pi(A,X)} \left( \frac{K}{K-1} R - X_*^\top B_* \right) X_* \right\}, \tag{14}$$

$$\frac{\partial^2 l_n(B_*)}{\partial B_*^2} = 2\mathbb{P}_n \left\{ \frac{1}{w(A,X)\pi(A,X)} X_* X_*^\top \right\}. \tag{15}$$

Assuming that $\hat{B}_* \xrightarrow{P} B_*$ and using Taylor Series approximations similar to Proposition 1,

$$\begin{aligned}
&\sqrt{n}\left(\hat{B}_* - B_*\right) \\
&= \sqrt{n} E \left\{ \frac{X_* X_*^\top}{w(A,X)\pi(A,X)} \right\}^{-1} \left\{ \mathbb{P}_n \frac{X_*}{w(A,X)\pi(A,X)} \left( \frac{K}{K-1} R - X_*^\top B_* \right) \right\} + o_p(1) \\
&= E \left\{ \frac{X_* X_*^\top}{w(A,X)\pi(A,X)} \right\}^{-1} \left\{ \sqrt{n}\mathbb{P}_n \frac{\epsilon X_*}{w(A,X)\pi(A,X)} \right\} + o_p(1), \text{ and}
\end{aligned} \tag{16}$$

$$\begin{aligned}
&var\left\{ \sqrt{n}\left(\hat{B}_* - B_*\right) \right\} \\
&= E \left\{ \frac{X_* X_*^\top}{w(A,X)\pi(A,X)} \right\}^{-1} E \left\{ \frac{X_* X_*^\top \sigma_0^2(A,X)}{w(A,X)^2 \pi(A,X)^2} \right\} E \left\{ \frac{X_* X_*^\top}{w(A,X)\pi(A,X)} \right\}^{-1} + o(1).
\end{aligned} \tag{17}$$

The format of (17) is now very similar to that of Proposition 1. Taking a functional derivative, as in the proof for Proposition 1, shows that the sandwich variance is minimized at $w(A,X) = \dfrac{\sigma_0^2(A,X)}{\pi(A,X)}$. Thus, the proof is complete.

## Web Appendix B

This section includes proofs of Lemma 1, Theorem 1, Theorem 2, and Theorem 3.

[**Proof of Lemma 1**] For the first case, linear regression, assume that $\Phi(A,X)$ is a finite-dimensional collection of features in $(A,X)$ and functions of the features. $\Phi(A,X)$ includes

main effects, and can additionally include squared and intercept terms, cubic terms, etc. Assume that $E\left\{\Phi(A, X)\Phi(A, X)^\top\right\}$ is bounded and positive definite and let $\hat{\sigma}_n^2(A, X)$ be a linear function predicting the squared residuals from the features in the set $\Phi(A, X)$. Then, the prediction function, $\hat{\sigma}_n^2(A, X)$, is uniformly consistent for $\sigma_0^2(A, X)$, and moreover is a vector process which belongs to a VC-class via Lemma 9.6 of Kosorok (2008). This implies that it is bounded uniform integrable (belongs to a BUEI class) and therefore in a Donsker class because of the measurability conferred by Lemma 8.12 of Kosorok (2008).

For the second case, it is known that random forest predictions can be characterized by $\hat{\sigma}_n^2(A, X) = \sum_{j=1}^\infty \alpha_j c_j \mathbb{I}\left\{(a, x) \in R_j\right\}$, where each $R_j$ is a hyperrectangle, $c_j$ is the expected value of $Y$ over the region $R_j$, and $\sum_{j=1}^\infty \alpha_j = 1$. The random forest prediction function, $\hat{\sigma}_n^2(A, X)$, is therefore a convex combination of expected values over hyperrectangular regions. Since the class of hyperrectangles of fixed dimensions is a VC-class and $\hat{\sigma}_n^2(A, X)$ is a convex combination of them, $\hat{\sigma}_n^2(A, X)$ belongs to a convex hull class, $\mathcal{C}$.

For a convex hull class, Corollary 9.5 of Kosorok (2008) applies and indicates that the covering number of $\mathcal{C}$ is bounded above by $K\left(\frac{1}{\epsilon}\right)^{2-2/V}$, where $V$ is the VC-index of $\mathcal{C}$. Taking the square root results in $\epsilon^{-\alpha}$ where $\alpha \leqslant 1$, resulting in a bounded integral when integrated. Therefore, $\hat{\sigma}_n^2(A, X)$ belongs to a BUEI class and can also be shown to be Pointwise Measurable (PM) (see, e.g., Section 8.2 of Kosorok (2008)).

We wish first to show that $\frac{1}{\hat{\sigma}_n^2(A, X)}$ also belongs to a class that is BUEI and PM. For $g(x) = \frac{1}{x}$, $\frac{\partial}{\partial x}g(x) = \frac{-1}{x^2} \leqslant \frac{-1}{c_1^2}$ since $x$ represents the quantity $\hat{\sigma}_n^2(A, X)$. Therefore, $g(x)$ is a Lipschitz continuous function, and Lemma 9.17 (vi) of Kosorok (2008) shows that $\frac{1}{\hat{\sigma}_n^2(A, X)}$ is BUEI and PM using $c_1 \leqslant \infty$ as an envelope. Finally, using Lemma 9.17 (v), $\frac{X\epsilon}{\hat{\sigma}_n^2(A, X)}$ also belongs to a class which is BUEI and PM, using $E\left(|X\epsilon|\right) \leqslant \|X\|\sigma_0$ as an envelope. Since $\frac{X\epsilon}{\hat{\sigma}_n^2(A, X)}$ belongs to a class which is BUEI and PM, by Theorem 8.19, the class is also P-Donsker.

[**Proof of Theorem 1**] Considering $f^{opt}(X) \in \mathcal{F}$, where $f^{opt}(X) = E\left\{ \frac{RA}{\pi(A, X)} \Big| X \right\} = 2\delta(X)$, let $Y^* = 2RA$, and characterize the estimated variance function, $\hat{\sigma}_n^2(A, X)$, as the matrix $\hat{\Sigma}_n \in \mathbb{R}^{n \times n}$ where $\hat{\Sigma}_n = \text{diag}(\hat{\sigma}_n^2(a_1, x_1), ..., \hat{\sigma}_n^2(a_n, x_n))$. The least squares SD-Learning estimate for $\beta_0$ can be written as:

$$
\begin{aligned}
\hat{\beta}_n^S &= (X^\top \hat{\Sigma}_n^{-1} X)^{-1} (X^\top \hat{\Sigma}_n^{-1} Y^*) \\
&= \beta_0 + \left( \frac{1}{n} X^\top \hat{\Sigma}_n^{-1} X \right)^{-1} \left( \frac{1}{n} X^\top \hat{\Sigma}_n^{-1} \epsilon \right)
\end{aligned}
\tag{18}
$$

We show that the second term in this equation converges to zero. Note that since $\sigma_0^2(A, X) \geqslant c_1$ by Assumption 1, and $\hat{\sigma}_n^2(A, X)$ is uniformly consistent for $\sigma_0^2(A, X)$ by Assumption 3, there exists $N$ for which $\hat{\sigma}_n^2(A, X) \geqslant \frac{c_1}{2}$ for all $n \geqslant N$. Letting $t \in \mathbb{R}^p$ be an arbitrary vector on the unit sphere,

$$
\begin{aligned}
\Bigg| \frac{1}{n} \sum_{i=1}^n & \frac{t^\top x_i x_i^\top t}{\hat{\sigma}_n^2(a_i, x_i)} - \frac{1}{n} \sum_{i=1}^n \frac{t^\top x_i x_i^\top t}{\sigma_0^2(a_i, x_i)} \Bigg| \\
&\leqslant \frac{1}{n} \sum_{i=1}^n \left( t^\top x_i \right)^2 \left| \frac{1}{\hat{\sigma}_n^2(a_i, x_i)} - \frac{1}{\sigma_0^2(a_i, x_i)} \right| \\
&\leqslant \frac{1}{n} \sum_{i=1}^n \| x_i \|^2 \cdot \frac{2}{c_1^2} \left\| \hat{\sigma}_n^2(A, X) - \sigma_0^2(A, X) \right\|_{\mathcal{A}, \mathcal{X}} + o_p(1) \\
&= \left\{ \frac{2}{c_1^2} \left\| \hat{\sigma}_n^2(A, X) - \sigma_0^2(A, X) \right\|_{\mathcal{A}, \mathcal{X}} \right\} \cdot \frac{1}{n} \sum_{i=1}^n \| x_i \|^2 + o_p(1) \\
&= o_p(1) \cdot O_p(1) + o_p(1), \text{ and}
\end{aligned}
\tag{19}
$$

$$
\begin{aligned}
\Bigg| \frac{1}{n} \sum_{i=1}^n & \frac{X_i \epsilon_i}{\hat{\sigma}_n^2(a_i, x_i)} - \frac{1}{n} \sum_{i=1}^n \frac{X_i \epsilon_i}{\sigma_0^2(a_i, x_i)} \Bigg| \\
&= \left| \frac{1}{n} \sum_{i=1}^n X_i \epsilon_i \left\{ \frac{1}{\hat{\sigma}_n^2(a_i, x_i)} - \frac{1}{\sigma_0^2(a_i, x_i)} \right\} \right| \\
&\leqslant \left| \frac{1}{n} \sum_{i=1}^n X_i \epsilon_i \left\{ \frac{2 \left\| \hat{\sigma}_n^2(A, X) - \sigma_0^2(X) \right\|_{\mathcal{A}, \mathcal{X}}}{c_1^2} \right\} \right| + o_p(1) \\
&= \frac{2 \left\| \hat{\sigma}_n^2(A, X) - \sigma_0^2(X) \right\|_{\mathcal{A}, \mathcal{X}}}{c_1^2} \cdot \left| \frac{1}{n} \sum_{i=1}^n X_i \epsilon_i \right| + o_p(1),
\end{aligned}
\tag{20}
$$

where $n^{-1} \sum_{i=1}^n \| x_i \|^2$ is $O_p(1)$ because it converges to $E \left( XX^\top \right)$ and $\left| \frac{1}{n} \sum_{i=1}^n X_i \epsilon_i \right|$ converges to $|E(X_i \epsilon_i)| = 0$. (19) shows that $n^{-1} X^\top \hat{\Sigma}_n^{-1} X \xrightarrow{P} n^{-1} X^\top \Sigma_0^{-1} X$, since it is true for arbitrary

$t$ on the unit sphere, and (20) shows that $n^{-1}X^\top\hat{\Sigma}_n^{-1}\epsilon \xrightarrow{P} n^{-1}X^\top\Sigma_0^{-1}\epsilon$. Since inversion is a continuous operation when the limit is a positive definite matrix, $\left(n^{-1}X^\top\hat{\Sigma}_n^{-1}X\right)^{-1} \xrightarrow{P} \left(n^{-1}X^\top\Sigma_0^{-1}X\right)^{-1}$ by the Continuous Mapping Theorem. Since $\left(n^{-1}X^\top\Sigma_0^{-1}X\right)^{-1}$ is finite and positive definite while $n^{-1}X^\top\Sigma_0^{-1}\epsilon$ converges to zero using the Kolmogorov Law of Large Numbers, $\left(n^{-1}X^\top\hat{\Sigma}_n^{-1}X\right)^{-1}\left(n^{-1}X^\top\hat{\Sigma}_n^{-1}\epsilon\right)$ converges to zero, and $\hat{\beta}_n^S \xrightarrow{P} \beta_0$.

[**Proof of Theorem 2**] We show that the preconditions are met for Lemma 4.4 of Kosorok (2008). Assumptions 1 and 2 guarantee that $U_0$ is positive definite. Additionally, $\hat{\beta}_n^S \xrightarrow{P} \beta_0$ based on Theorem 1. $\mathbb{P}_n\left\{\dfrac{XX'}{\hat{\sigma}_n^2(A,X)} - \dfrac{XX'}{\sigma_0^2(A,X)}\right\} \to 0$ was shown by (19). Now, consider the following:

$$
\begin{aligned}
\left|\frac{1}{n}\sum_{i=1}^n \left[t^\top x_i \epsilon_i\left\{\frac{1}{\hat{\sigma}_n^2(a_i,x_i)} - \frac{1}{\sigma_0^2(a_i,x_i)}\right\}\right]^2\right| & \\
&= \frac{1}{n}\sum_{i=1}^n (t^\top x_i)^2 \sigma_0^2(a_i,x_i)\left\{\frac{1}{\hat{\sigma}_n^2(a_i,x_i)} - \frac{1}{\sigma_0^2(a_i,x_i)}\right\}^2 + o_p(1) \\
&\leqslant \frac{c_2}{n}\sum_{i=1}^n (t^\top x_i)^2\left\{\frac{1}{\hat{\sigma}_n^2(a_i,x_i)} - \frac{1}{\sigma_0^2(a_i,x_i)}\right\}^2 \\
&\leqslant \frac{4c_2}{c_1^4}\cdot\frac{1}{n}\sum_{i=1}^n (t^\top x_i)^2\left\|\hat{\sigma}_n^2(a_i,x_i) - \sigma_0^2(a_i,x_i)\right\|_{\mathcal{A},\mathcal{X}}^2 + o_p(1) \\
&\leqslant \frac{4c_2}{c_1^4}\left\|\hat{\sigma}_n^2(A,X) - \sigma_0^2(A,X)\right\|_{\mathcal{A},\mathcal{X}}^2 \cdot\frac{1}{n}\sum_{i=1}^n \|x_i\|^2 + o_p(1) \\
&= o_p(1)\cdot O_p(1) + o_p(1).
\end{aligned}
\tag{21}
$$

Therefore, $\mathbb{P}_n\left\{\dfrac{X\epsilon}{\hat{\sigma}_n^2(A,X)} - \dfrac{X\epsilon}{\sigma_0^2(A,X)}\right\}^2 \xrightarrow{P} 0$, and the conditions for the Lemma 4.4 of Kosorok (2008) are met. Note the following two facts:

$$
\begin{aligned}
\hat{\beta}_n^S &= \mathbb{P}_n\left\{\frac{XX'}{\hat{\sigma}_n^2(A,X)}\right\}^{-1}\mathbb{P}_n\left\{\frac{XY^*}{\hat{\sigma}_n^2(A,X)}\right\} \\
&\implies \sqrt{n}\left(\hat{\beta}_n^S - \beta_0\right) = \left[\mathbb{P}_n\left\{\frac{XX'}{\hat{\sigma}_n^2(A,X)}\right\}^{-1}\right]\cdot\sqrt{n}\mathbb{P}_n\left\{\frac{X\epsilon}{\hat{\sigma}_n^2(A,X)}\right\}, \text{ and}
\end{aligned}
\tag{22}
$$

$$
\begin{aligned}
\sqrt{n}\left[\mathbb{P}_n\left\{\frac{X\epsilon}{\hat{\sigma}_n^2(A,X)}\right\} - E\left\{\frac{X\epsilon}{\sigma_0^2(A,X)}\right\}\right] &\to N\left[0, E\left\{\frac{XX'}{\sigma_0^2(A,X)}\right\}\right] \\
&\implies \sqrt{n}\mathbb{P}_n\left\{\frac{X\epsilon}{\hat{\sigma}_n^2(A,X)}\right\} \to N\left[0, E\left\{\frac{XX'}{\sigma_0^2(A,X)}\right\}\right],
\end{aligned}
\tag{23}
$$

where $E\left\{\dfrac{X\epsilon}{\sigma_0^2(A,X)}\right\} = 0$ in (23) stems from Assumption 4. With the combination of (22) and (23), we can establish:

$$\sqrt{n}\left(\hat{\beta}_n^S - \beta_0\right) \to E\left\{\frac{XX'}{\sigma_0^2(A,X)}\right\}^{-1} \cdot N\left[0, E\left\{\frac{XX'}{\sigma_0^2(A,X)}\right\}\right]$$

$$\to N\left[0, E\left\{\frac{XX'}{\sigma_0^2(A,X)}\right\}^{-1}\right]. \tag{24}$$

[**Proof of Theorem 3**] Since the SD-Learning framework is unified between the binary and multi-arm case, this proof is identical to the proofs of Theorems 1 and 2, with $X_* \in \mathbb{R}^{p(K-1)}$ instead of $X \in \mathbb{R}^p$ and $B_* \in \mathbb{R}^{p(K-1)}$ instead of $\beta_0 \in \mathbb{R}^p$.



\end{document}